\documentclass[aps,prl,10pt]{revtex4-1}
\usepackage{graphicx}
\usepackage{color}
\usepackage{natbib}

\begin{document}
\title{The quantum plasma lens concept: a preliminary investigation}
\author{F. Tanjia$^1$, R. Fedele$^1$, S. De Nicola$^{1,2}$, D. Jovanovic$^3$, A. Mannan$^4$}
\affiliation{{\small $^1$ Dipartimento di Fisica, Universi$\grave{a}$ di Napoli ``Federico II" and INFN, Napoli, Italy\\
$^2$ CNR-SPIN, Complesso Universitario  di Monte S'Angelo - Via Cintia, I-80126, Napoli, Italy\\ $^3$ Institute of Physics Belgrade, Serbia\\ $^4$ Dipartimento di Scienze Ambientali, Seconda Universit$\grave{a}$ degli Studi di Napoli, Caserta and INFN Napoli, Italy}}

\begin{abstract}
Recently, a theoretical investigation of the collective and nonlocal quantum effects has been carried out within the framework of a quantum approach to the relativistic charged particle beam travelling in a cold, collisionless, strongly magnetized plasma. This has been done taking into account both the plasma wake field excitation and the quantum paraxial approximation. On the basis of this theory, here we carry out a preliminary study of the transverse effects experienced by a cold relativistic beam through a thin plasma slab (plasma lens). In the strongly nonlocal regime, in which the beam experiences a very strong focusing effect, the scheme of plasma lens is reviewed in terms of the wave description provided by the above quantum theory.
\end{abstract}

\maketitle

\section{Introduction \label{Introduction}}
The plasma lens is a thin plasma slab that is used to focus charged particle beams as well as laser pulses. Due to the electromechanical actions that electromagnetic (e.m.) or charged particle beams can experience, the plasma can produce acceleration or focusing on a beam traveling through it. By making use of a thin plasma lens, it is possible to compress and focus a laser or charged particle beam outside of the plasma in the vacuum region. Historically, plasmas had been used to focus a continuous low-energy electron beam by beam-ionized gas within a cathode ray tube way back in 1922 \cite{Johnson1922}. In 1987 Chen first proposed the concept that plasma lenses could form an immensely strong final focusing system for a linear collider \cite{Chen1987a}. At that time, the analog focusing of e.m. radiation beams by means of a plasma was already well developed. However, it became more popular just when the first proposals of large amplitude plasma wave excitation driven by e.m. wavepackets or very short e.m. pulses with a transverse profile appeared in literature.  Then, the compression and focusing of relativistic charged particle beams or high e.m. intensity laser beams through Laser-driven \cite{Rosenbluth1972,Tajima1979,Gorbunov1987b,Sprangle1988} or charged particle beam-driven \cite{Chen1985, Rosenzweig1988, Rosenzweig1991} excitations have been of great interest during the last two decades.

The typical charged particle beam-driven plasma wave excitation is the well known Plasma Wake Field (PWF) excitation. In the PWF excitation, a charged particle beam or bunch (driver) travels in a neutral plasma with ions as background.\newline
If the driver is a relativistic electron beam, then the Coulomb force of the beam's space charge expels plasma electrons, which rush back in after the beam and produce a large amplitude plasma wave behind the driver. This plasma wave oscillates at the electron plasma frequency and follows the driver much the same way water wakes follow a fast boat (plasma wake) \cite{Chen1985}. Its phase velocity is therefore equal to the driver velocity, and is almost independent of the plasma density. The electromagnetic field associated to the wake (wake field) has transverse as well as longitudinal  components. Thus, a test particle experiences the effects of both the transverse (focusing/defocusing) and the longitudinal (acceleration/deceleration) components of the wake field. Depending on the regimes, the test particle can be the one of a secondary beam externally injected in phase locking with the wake (driven beam) or belonging to the driver. In the last circumstance, the driver experiences the effects of the wake field that itself produced. Taking into account all together effects on each particle of the driver we can describe the collective self-interaction of the driver with the plasma.\newline
In the case of positrons, electrons of the plasma background are pulled in by the driver which overshoot and set up the plasma oscillation. Then, it is easily seen that we can provide PWF interaction for a relativistic positron beam in a way fully similar to the one described above for an electron beam.

In PWF-based plasma lens, when a relativistic electron (positron) beam passes through a plasma slab with a thickness of few of $c/\omega_p$ ($\omega_p$ is the plasma frequency), since the time of beam-plasma interaction is rather small, the radial component of the wake fields produces a bending of the trajectories inward to the propagation axis. Once the particles are again in vacuum, the wake fields are off. The repulsive electric force between the particles due to the beam space charge is initially almost balanced by the attractive magnetic force and then they will converge to a minimum spot (focus) \cite{Chen1987a, Chen1987b, Su1990, Fedele1990}.

Several analytical, numerical, as well as experimental investigations have been carried out to describe the behavior of the plasma lens characterized in the two regimes depending on the ratio of plasma density $n_p$ to beam density $n_b$. In an overdense plasma lens where $n_p \gg n_b$, the space charge of the electron beam is fully neutralized by the plasma through the displacement of plasma electrons by the beam electrons, resulting in beam self-focusing through its own magnetic field \cite{Rosenzweig1990, Nakanishi1991, Hairapetian1994, Hairapetian1995, Govil1999}. In the underdense lens, where $n_p\lesssim n_b$, all plasma electrons are displaced by the beam electrons, and the focusing force is due to the remaining plasma ions \cite{Su1990,Barov1994}.

With a plasma slab of very small thickness, the so called \textit{thin plasma lens approximation} \cite{Chen1987a,Su1990,Fedele1990} corresponds to a \textit{kick approximation}, the beam wave function remains almost unchanged except for the appearance of a finite value of a chirping phase factor. Thus the transverse beam spot size remains almost unchanged inside the plasma lens. For instance, a beam with an initial $\sigma_0=\sigma(z=0)$ and $\sigma_0' \equiv (d\sigma/dz)_{z=0} = 0$ [$\sigma(z)$ being the transverse beam spotsize] enters a thin plasma slab of thickness $l$, according to the \textit{thin plasma lens approximation}

$$\sqrt{K}l\ll 1,$$
where $K$ is the focusing strength. In this approximation, the focal length of the lens becomes
\begin{equation}\label{p1}
f=|\rho(z=l)|=\frac{1}{Kl}\left(1-\frac{\epsilon_t^2}{K\sigma_0^4}\right)^{-1},
\end{equation}
where $\rho(z)=\sigma(z)/\sigma'(z)$ can be interpreted as the bending radius of the surface normal to envelope trajectory that intersects the longitudinal axis at location $z$ and $\epsilon_t$ is the transverse thermal emittance of the beam. Typically, in a strong plasma lens the focusing effect dominates the emittance spreading, i.e., $\epsilon_t^2/K\sigma_0^4\ll 1$ is satisfied. Thus, with this approximation, Eq. (\ref{p1}) becomes
\begin{equation}\label{p2}
f\simeq\frac{1}{Kl}.
\end{equation}
\begin{figure}
  \centerline{\includegraphics[width=12cm]{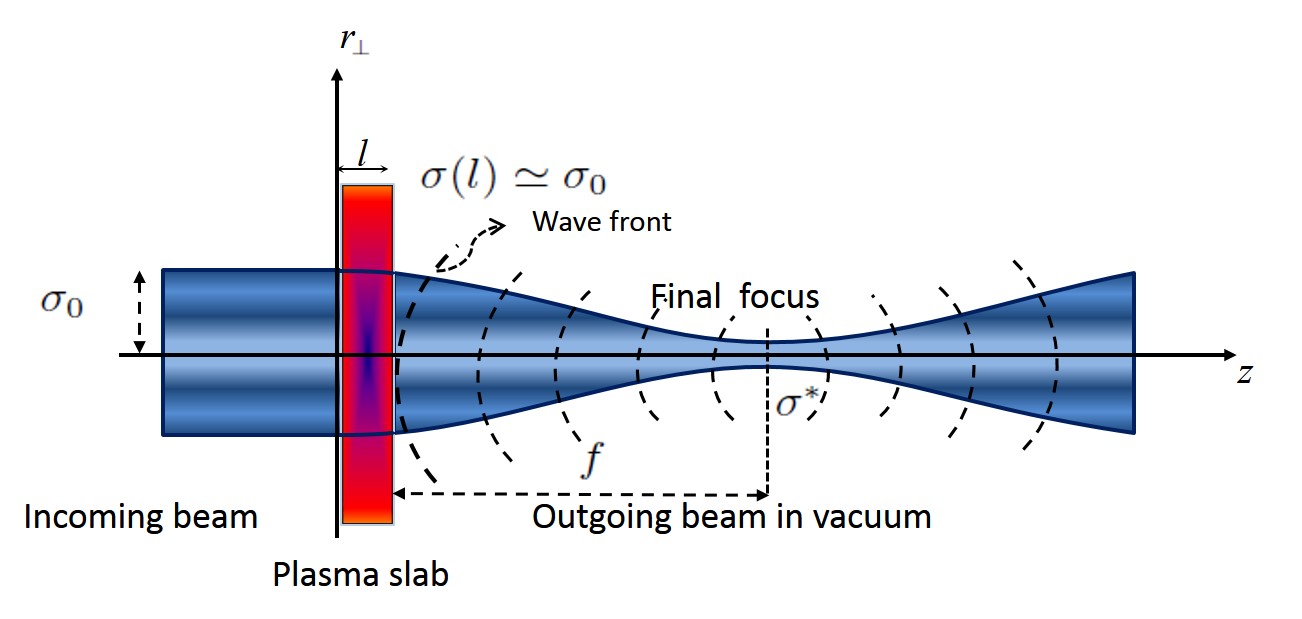}}
  \caption{Scehmatic diagram of plasma lens}\label{f1-1}
\end{figure}
In this paper, on the basis of recently developed theory \cite{Fedele2012a,Fedele2012b}, we propose a plasma lens to describe the transverse effect on a cold relativistic electron/positron beam traveling along $z$-axis through a magnetized thin plasma slab. We adopt an effective wave description that associates a macroscopic wave function to  the beam. We describe the beam wave function $\psi$ in terms of the beam density $\rho_b$, i.e., $\psi=\sqrt{\rho_b}~e^{i\theta}$, where $\theta$ is the eikonal. According to the picture of surface normal to the envelope trajectories given above, $\rho(z)$ is the bending radius, at each $z$, of the the locus of the points of constant phase $\theta$ provided by the beam wave function. A preliminary analysis to investigate the transverse effect of the beam spot size to reach the nano scales in the final focus is carried out in the aberration-less approximation. The manuscript is organized as follows. In section \ref{equation}, the beam-plasma model in the quantum regime in the presence of an external uniform magnetic field is presented in terms of a pair of coupled partial differential equations. One is a sort of nonlinear Schr\"odinger equation for $\psi$ whose nonlinear term is proportional to the wake potential and, in turn, satisfies a Poisson-like equation. This system is then specialized to the case of strong focusing and the corresponding plasma lens theory is presented under the aberration-less approximation. The numerical analysis of this thin lens is carried out in section \ref{numerical}. It is shown that the beam reaches the nano scale at the interaction point (final focus). Finally, the conclusion and remarks are given in section \ref{conclusion}.

\section{Governing equations} \label{equation}
Recently, we developed a theory \cite{Fedele2012a,Fedele2012b}, where the quantum transverse beam motion for a cold relativistic charged particle beam traveling in a cold, collisionless, strongly magnetized plasma is carried out. A Lorentz-Maxwell fluid model of the ``Beam + Plasma" system is adopted in the presence of a constant and uniform ambient magnetic field in the overdense regime. In the long beam limit, by using the Hartree's mean field approximation \cite{Hartree1957}, the individual quantum nature of the beam particles is taken into account via PWF excitation (quantum paraxial diffraction). We consider typical range of densities, energies and temperatures of the particle such that the beam is sufficiently cold to take into account the individual quantum nature of the particles (single-particle uncertainty principle and spin), but sufficiently warm to disregard the collective nature of the beam particles due to overlapping of the wave function (exchange effects). On the basis of this theory, the self-consistent beam-plasma interaction that accounts for the PWF excitations and the quantum paraxial beam diffraction, in cylindrical symmetry,  is governed by the following system of equations \cite{Fedele2012a,Fedele2012b}
\begin{eqnarray}
&&i\epsilon_c\frac{\partial \psi_m}{\partial z}=
-\frac{\epsilon_c^2}{2}\frac{1}{r_\perp}\frac{\partial}{\partial
r_\perp}\left(r_\perp\frac{\partial \psi_m}{\partial
r_\perp}\right)+U_w[|\psi_m|^2]\psi_m
+\left(\frac{1}{2}Kr^2_\perp+\frac{m^2\epsilon_c^2}{2r_\perp^2}\right)\psi_m
\label{c3}\\
&&\frac{1}{r_\perp}\frac{\partial}{\partial
r_\perp}\left(r_\perp\frac{\partial U_w}{\partial
r_\perp}\right)-\frac{k^4_{pe}}{k_{uh}^2}\,U_w=
\frac{k^4_{pe}}{k_{uh}^2}
\frac{N}{n_0\gamma_0\sigma_z}\,|\psi_m|^2, \label{c4}\
\end{eqnarray}
where $k_{pe}=\omega_{pe}/c$, $k_{uh}=\omega_{uh}/c$, $\omega_{pe}=\left(4\pi n_0 e^2/m_0\right)^{1/2}$ is the {electron plasma} frequency, $\omega_{uh}=\left(\omega_{pe}^2+\omega_{ce}^2\right)^{1/2}$ is the upper hybrid frequency, $\omega_{ce}=-e B_0/m_0 c$ is the electron cyclotron frequency, $K=\omega_{ce}^2/4\gamma_0^2 c^2$,
$\epsilon_c\equiv\hbar/m_0\gamma_0c=\lambda_c/\gamma_0$, $n_0$ is the electron plasma density, $e$ is the absolute value of an electron charge, $m_0$ is the rest mass of
the electron/positron, $B_0$ is the constant and uniform external magnetic field acting along the same direction as the propagation direction of the beam, viz., $\mathbf{B}_0 = B_0\hat{z}$, $U_w(r_\perp,z)$ is the dimensionless wake potential, $N$ is the total number of particles, $\sigma_z$ is the beam length, $\gamma_0$ is the relativistic factor, and $m$ is an integer representing the vortex charge to describe the effect of vorticity due to the orbital angular momentum. Note that, $\lambda_c$ is the Compton wavelength and, therefore, the quantity $\epsilon_c$ plays the role of the \textit{relativistic Compton wavelength}. We would also like to point out the fact that the origin of $\epsilon_c$ described here, is completely different from that of the earlier works (defined as $\epsilon_t$ in previous section) of plasma lens \cite{Chen1987a,Su1990,Fedele1990}. In all the previous works it has a thermal origin (thermal spreading of the electron rays) and taken into in the envelope equation (rms beam description) through the beam emittance \cite{Lawson1988}. The single-particle fundamental emittance $\epsilon_c$ described in this theory has a quantum origin by taking into account the individual quantum nature of the beam particles and disregarding the overlapping of the wave function.

Note that, with respect to the refs. \cite{Fedele2012a,Fedele2012b}, we replaced for simplicity $\xi$ with $z$ that is now the time-like coordinate in the co-moving frame. This dynamics is governed by the interplay between the attraction and repulsion among the particles due to the magnetic and electric fields that are produced within the beam itself. In particular, when in Eq. (\ref{c4}) we can take the approximation $\nabla_\perp^2U_w\gg \left(k_{pe}^4/k_{uh}^2\right)U_w$, the attraction among the beam particles takes place in such a way that the beam may exhibit a strong self-focusing \cite{Jovanovic2012b,Jovanovic2013}. On the other hand, when the beam travels \textit{in vacuo}, the repulsion among the particles due to space charge electric field exceeds, to the order of $1/\gamma_0^2$, the magnetic attraction. In this regime, the effective potential experienced by the beam is then governed by a Poisson equation whose source term is given by the beam density.

By assuming that $m$ is fixed to 0, we investigate the transverse dynamics of an electron/positron beam, whose initial distribution is purely Gaussian. We also assume that the beam is traveling  through a plasma slab (plasma lens), along the external magnetic field $\mathbf{B}_0$ and subsequently in a vacuum (where the external magnetic field is present as well). By adopting the above mentioned quantum model of charged particle beam \cite{Fedele2012a,Fedele2012b}, we study the physical conditions that allow the beam to experience a strong focusing outside the plasma lens. The latter, indeed, considered here, is sufficiently thin to provide a strong transverse kick to the beam particles that changes suddenly their momentum distribution but leaves their space distribution practically unchanged within the lens.

According to all the assumptions presented above, the wave-like description of the electron/positron transport within a lens of length $l$ and subsequently \textit{in vacuo} is provided by the following pair of governing equations

\begin{eqnarray}
&&i\epsilon_c\frac{\partial \psi}{\partial z}=-\frac{\epsilon_c^2}{2}\frac{1}{r_\perp}\frac{\partial}{\partial r_\perp}\left(r_\perp\frac{\partial \psi}{\partial r_\perp}\right)+U\left[\,|\psi|^2\,\right]\psi+\frac{1}{2}Kr^2_\perp\psi\label{b7}\\
&&\frac{1}{r_\perp}\frac{\partial}{\partial r_\perp}\left(r_\perp\frac{\partial U}{\partial r_\perp}\right)=\alpha\,|\psi|^2, \label{b8}\
\end{eqnarray}

where $$\alpha=\left\{
  \begin{array}{ll}
    \left(k_{pe}^4/k_{uh}^2\right)\left(N/2\pi n_0\gamma_0\sigma_z\right)\equiv\alpha_w\,,\,\,\,\, \mbox{for}\,\,0\leq z \leq l\\
     \\
    -2q^2N/\sigma_z m_0\gamma_0^3c^2\equiv\alpha_{sp}\,,\,\,\,\, \mbox{for}\,\,z > l
  \end{array}
\right.$$

\subsection{Aberration-less approximation}
Note that the strong local regime we have assumed is compatible with the physical condition of strong beam focusing. Then, according to Ref. \cite{Fedele2013}, we can assume that the beam density is sufficiently picked around the propagation direction, that $U(r_\perp,z)$ can be expanded in power of $r_\perp$, around $r_\perp=0$, up to the second one. Therefore,
\begin{equation}
U=\frac{1}{2}\frac{\alpha}{\sigma^2(z)}\,r_\perp^2\,, \label{u1}\
\end{equation}
where $\sigma (z)$ is the beam spot size. In fact, the collective effects that $U(r_\perp,z)$ depends on the beam scale. A complete set of eigenfunctions of the system of equations (\ref{b7}) and (\ref{b8}) consistent with eq. (\ref{u1}) is given by the Laguerre-Gauss solutions. For sake of simplicity, among them we take the fundamental mode (aberration-less solution), i.e.
\begin{eqnarray}
&&\psi(r_\perp,z)=\frac{1}{\sqrt{\pi}\sigma_c(z)}\exp\left(-\frac{r_\perp^2}
{2\sigma_c^2(z)}+\frac{ir_\perp^2}{2\epsilon_c\rho_c(z)}\right)
\exp\left[i\phi_c(z)\right]\,, \label{d2}\
\end{eqnarray}
where $\sigma_c$ is the transverse rms of the quantum single-particle wave function (i.e., single quantum electron ray). Note that $\sigma_c(z)\ll\sigma(z)$. Furthermore, it is easily seen that the quantities $\sigma_c$, $\rho_c$ and $\phi_c$ satisfy the following set of coupled ordinary differential equations, viz.
\begin{eqnarray}
&&\frac{1}{\rho_c}=\frac{1}{\sigma_c}\frac{d\sigma_c}{dz}, \label{d3a}\\
&&\frac{d\phi_c}{dz}=-\frac{\epsilon_c}{\sigma_c^2}, \label{d3b}\\
&&\frac{d^2\sigma_c}{dz^2}+\overline{K}\sigma_c-\frac{\epsilon^2}{\sigma_c^3}=0, \label{d5a}\
\end{eqnarray}
where $\overline{K}(z)=K+\alpha/\sigma^2(z)$.  According to Ref.\cite{Fedele2013}, the evolution of the quantum single-particle spot size contains the same information carried out by an evolution equation for the beam spot size (description of the beam as a whole constituted by Fermions). Such an evolution equation exhibits, through the collective interaction, an amplification of the single-particle quantum effects. In fact, the evolution equation for the single-particle spot size, say $\sigma_c$, contains the fundamental emittance $\epsilon_c$. Then, the assumption $\sigma\approx\sqrt{N_\perp}\sigma_c$ and $\epsilon\approx N_\perp\epsilon_c$ define transverse beam spot size and emittance of the lowest quantum state compatible with the condition of \textit{non overlapping} of the Fermions. Here $N_\perp$ is the number of quantum electron rays that are intersecting the transverse plane (i.e., number of particles distributed in a single transverse plane). Therefore, we easily obtain (Sacherer equation)
\begin{equation}\label{Sacherer-eq-1}
\frac{d^2\sigma}{dz^2}+K\sigma + \frac{\alpha}{\sigma}-\frac{\epsilon^2}{\sigma^3}=0\,.
\end{equation}
Note that this macroscopic equation, written for the beam spot size at the lowest quantum state, is fully equivalent to the one for the quantum single-particle averaged motion in the presence of collective effects (for details, see Ref. \cite{Fedele2013}). This provides a macroscopic manifestation of the single-particle quantum effects.

\subsection{inside the slab}
We consider that the beam enters a thin plasma slab (plasma lens) of thickness $l$. According to both the conventional and the quantum theories, i.e., \cite{Chen1987a,Su1990,Fedele1990} as well as wave description \cite{Fedele2013}, the beam spot size remains almost unchanged inside the slab, i.e., $\sigma(l)\simeq \sigma_0$, where $\sigma_0$ is the initial (while entering the plasma) transverse beam spot size. Consequently, the focusing strength becomes almost constant, i.e., $\overline{K}=\overline{K}(\sigma_0)\simeq K+\alpha_w/\sigma_0^2$. Thus, with the approximation of thin lens, i.e., $\sqrt{\overline{K}}l\ll 1$ and strong focusing case, i.e., $\epsilon^2/\overline{K}\sigma_0^4\ll 1$, the focal length is
\begin{equation}
f=|\rho(z=l)|\simeq\frac{1}{\overline{K}l}. \label{t1}\
\end{equation}
\subsection{In vacuo}
Now, specializing the expansion (\ref{u1}) for the case of motion in vacuo, the effective focusing strength is now $\overline{K}=K+\alpha_{sp}/\sigma^2(z)$, so that eq. (\ref{d5a}) becomes
\begin{equation}
\frac{d^2\sigma}{dz^2}+K\sigma+\frac{\alpha_{sp}}{\sigma}-\frac
{\epsilon^2}{\sigma^3}=0. \label{c2}\
\end{equation}
Note that during the evolution in vacuo the effective focusing strength $\overline{K}$ is no longer constant as inside the lens. Note also, since $\alpha_{sp}$ is a negative quantity, there is an interplay among the focusing term $K\sigma$, the space charge blow up $\alpha_{sp}/\sigma$ and the beam spreading term $\epsilon^2/\sigma^3$. As $\sigma$ becomes smaller and smaller, during the process of focusing, the last two terms become greater and greater. For suitable choices of the parameters, one may expect that $\sigma$ reaches a minimum value for some $z$.
\begin{figure}
 \centerline{\includegraphics[width=6cm]{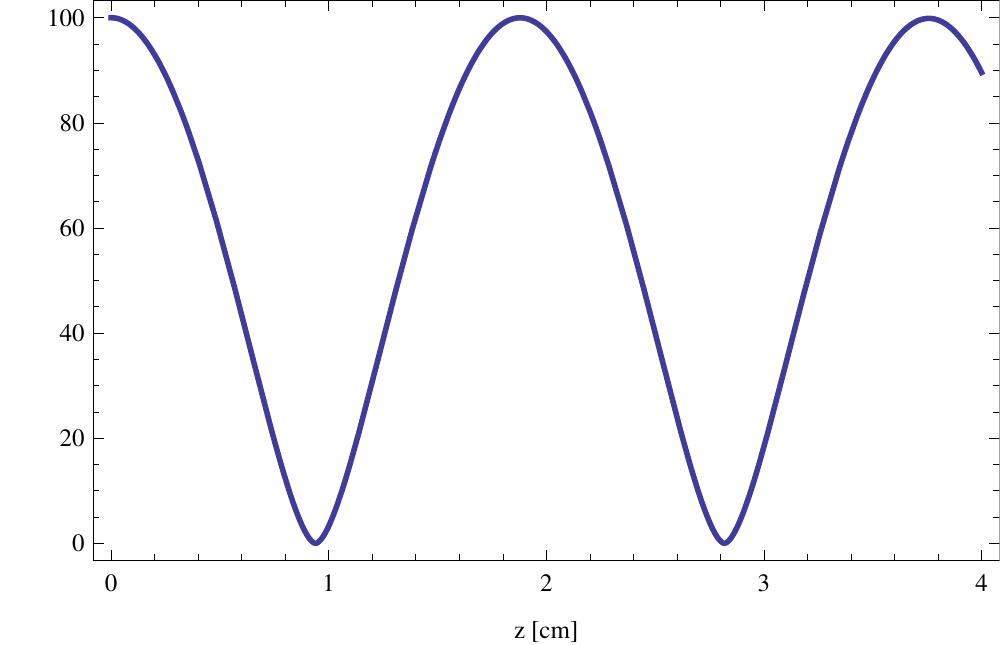}}
  \caption{Periodic Breathing of  $\sigma^2$ [in $\mu$m$^2$] as a function of $z$ [in cm] inside the plasma with the initial condition of $\sigma(0)=\sigma_0$ and $\sigma'(0)=0$.}\label{f1}
\end{figure}
\begin{figure}
  \centerline{\includegraphics[width=12cm]{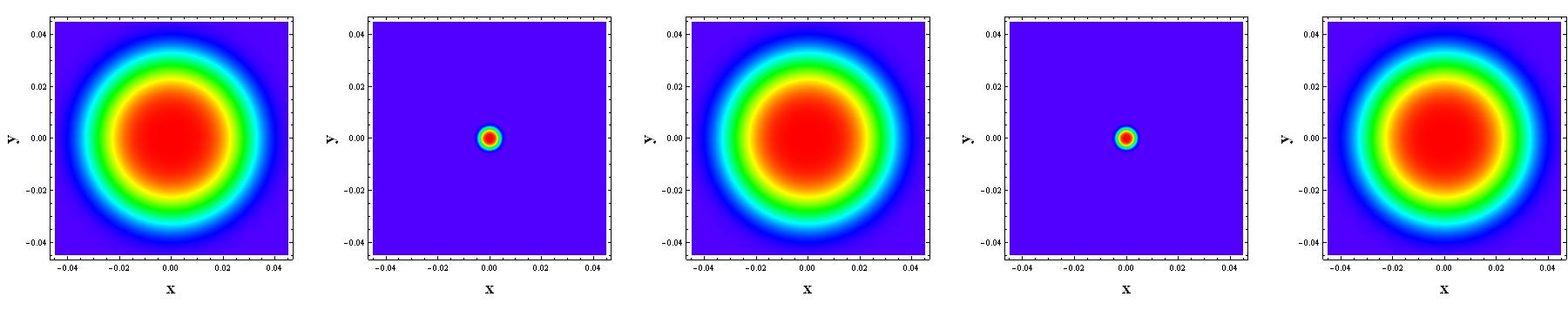}}
  \caption{Density plot of $|\psi^2|$ in transverse $x - y$ plane corresponding to FIG. 1. The focusing and defocusing of the profile indicates the periodicity of $\sigma$ inside the plasma }\label{f2}
\end{figure}
\begin{figure}
  \centerline{\includegraphics[width=6cm]{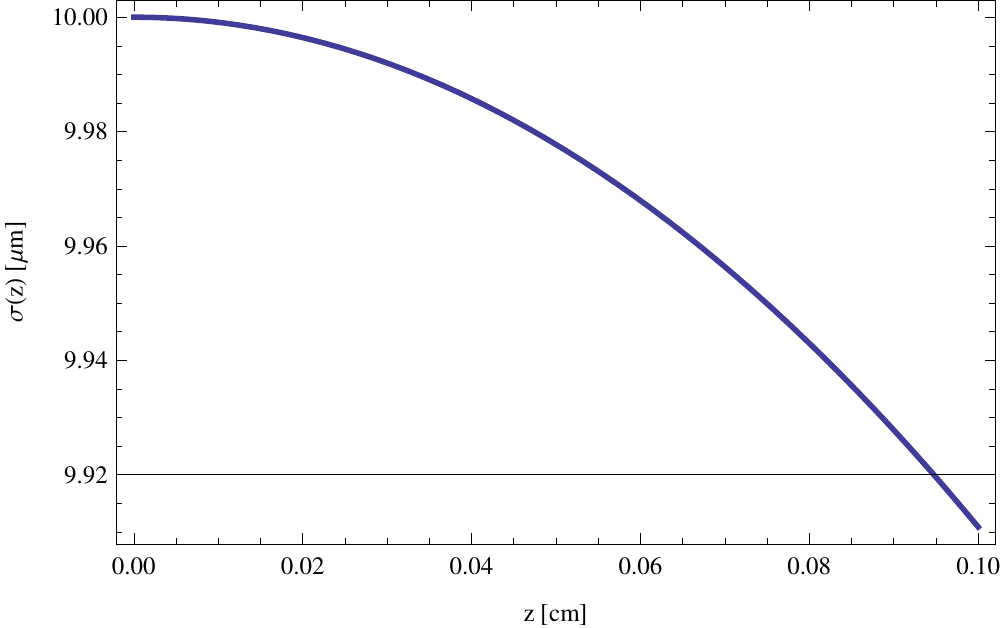}}
  \caption{Variation of $\sigma$ [in $\mu$m] as a function of $z$ [in cm] inside the lens with the initial condition of $\sigma(0)=\sigma_0$ and $\sigma'(0)=0$. For a lens of thickness $l=1$ mm, it has been found that $\sigma(l)=9.91\,\mu$m and $f=5.55$ cm.}\label{f3}
\end{figure}
\begin{figure}
   \centerline{\includegraphics[width=6cm]{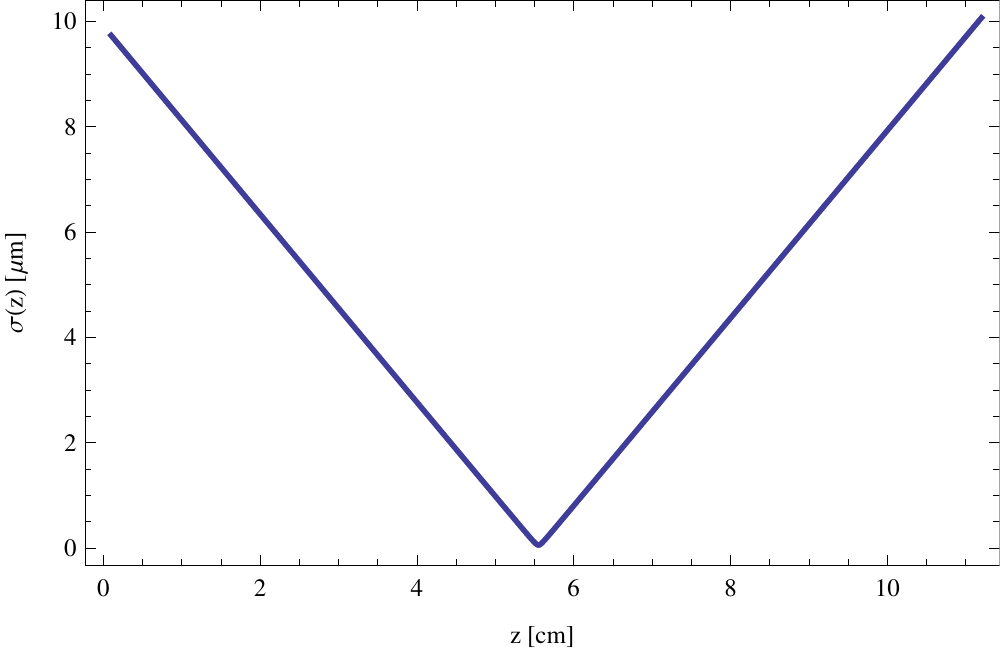}}
  \caption{Variation of $\sigma$ as a function of $z$ in the vacuum with initial condition $\sigma(0) = \sigma(l)$ and $\sigma'(0) = \sigma(l)/ \rho(l)$. $\sigma$ experiences strong focusing due to the dominance of the magnetic field term $K$. When $\sigma$ becomes very small near the final focus, the space charge effects becomes dominant compared magnetic field and it blows up. The beam spot size in the interaction point is $\sigma* = 56$ nm.}\label{f4}
\end{figure}
\begin{figure}
   \centerline{\includegraphics[width=6cm]{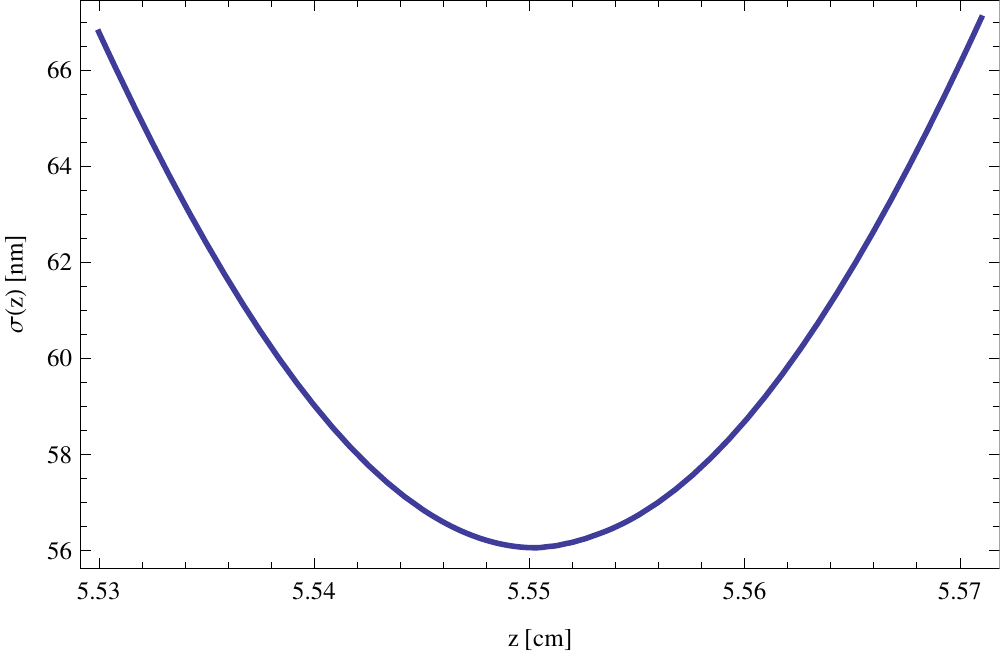}}
  \caption{A closer look on the variation of $\sigma$ around the interaction point. It shows the hyperbolic structure of $\sigma$}\label{f5}
\end{figure}
\begin{figure}
   \centerline{\includegraphics[width=6cm]{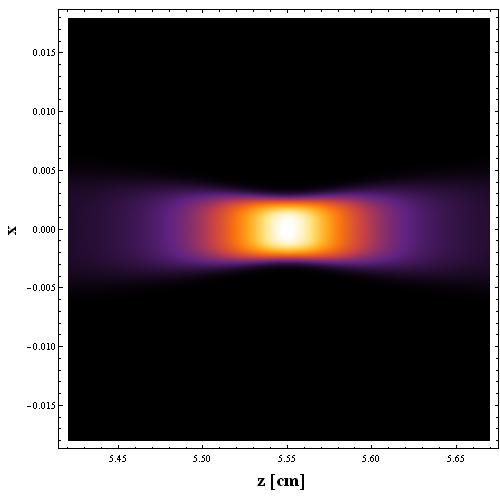}}
  \caption{The cross section of $\sigma$ with continuous $z$ around the interaction point corresponding to Figure \ref{f5}, showing the hyperbolic shape}\label{f6}
\end{figure}
\begin{figure}
   \centerline{\includegraphics[width=5cm]{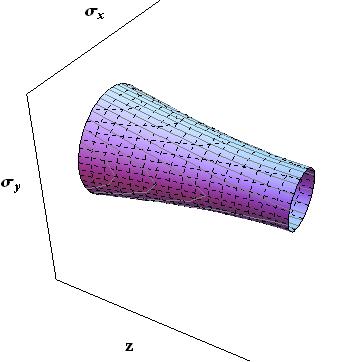}}
  \caption{The focusing of $\sigma$ with continuous $z$ till the interaction point shows the trumpet structure}\label{f7}
\end{figure}
\begin{figure}
   \centerline{\includegraphics[width=12cm]{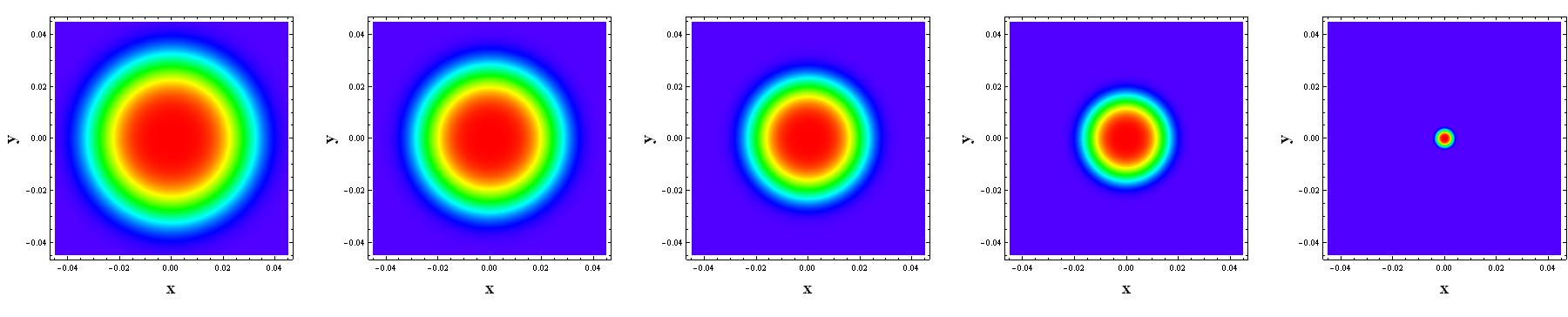}}
  \caption{Density plot of $|\psi|^2$ in the transverse $x - y$ plane at the final focus corresponding to Figures \ref{f5} and \ref{f6}}\label{f8}
\end{figure}

\section{Numerical evaluation}\label{numerical}
According to \cite{Fedele2012a,Fedele2012b}, we consider the quantum regime such that the overlapping of the single-particle wave functions is absent and the quantum paraxial  beam behavior takes into account the individual quantum nature of the particles (uncertainty principle and spin). This implies that the inter-particle distance is much greater than the thermal de Broglie wavelength. For instance, this condition is attainable for a typical range of the beam densities $n_b\sim10^{13} - 10^{16}$ cm$^{-3}$, energies $\gamma_0\sim10^2 - 10^4$, and relativistic transverse temperature $T_\perp\sim 10^2 - 10^3$ K. Thus, we can show that in the region of the interaction point (final focusing stage of a linear collider), our beam can reach spotsizes that are roughly ranging from a few 0.1 to a few 10 nm \textit{(nanobeams)}. We chose $n_0=10^{17}$ cm$^{-3}$, $n_b=10^{16}$ cm$^{-3}$, $\sigma_0=10$ $\mu$m, $\sigma_z=1$ mm, $\epsilon=10^{-9}$cm rad, $\gamma_0=10^4$, and $B_0=10^4$ Gauss.

\subsection{Inside the lens}
We numerically solved Eq. (\ref{Sacherer-eq-1}) (for $\alpha=\alpha_w$) for the above set of parameters with the initial condition of $\sigma(0)=\sigma_0$ and $\sigma'(0)=0$. For a thick plasma as medium, the periodic transverse breathing of $\sigma^2$ as a function $z$ is observed with a maximum value of $\sigma_0$ shown in Figure \ref{f1}. The corresponding density plot of $|\psi|^2$ also represents the periodic focusing and defocusing with $z$ as shown in Figure \ref{f2}. When the thick plasma medium is replaced by a slab of thickness $l=1$mm, the periodic breathing of $\sigma$ is no longer existing, rather it shows a slight bending of few fractions of $\mu$m keeping the value of $\sigma$ almost unchanged inside the slab, as shown in Figure \ref{f3}. For a lens of thickness $l=1$ mm, we numerically found that $\sigma(l)=9.91 \mu$m, which is almost the same as $\sigma_0$ and in accordance with the prediction of thin plasma lens theory. The focal length is calculated numerically as $f=5.55$ cm, which is also in accordance with the theoretical approximation $f\simeq 1/\overline{K}l$.

\subsection{In vacuo}
When the beam leaves the lens and goes into the vacuum, $\sigma$ is now governed by Eq. (\ref{c2}). We numerically solved Eq. (\ref{c2}) for the same set of parameters with the initial condition that we get at the end of the plasma slab, i.e., $\sigma(0)=\sigma(l)$ and $\sigma'(0)=\sigma'(l)=\sigma(l)/\rho(l)$. A strong focusing of $\sigma$ until $f$ is observed due to the dominance of the magnetic field term $K\sigma$ over the space charge term $\alpha/\sigma$, as shown in Figure \ref{f4}. When $\sigma$ becomes much smaller reaching $f$, the term $\alpha/\sigma$ becomes dominant over $K\sigma$, and it blows up afterwards. A closer look on the variation of $\sigma$ around the interaction point and the corresponding cross section with continuous $z$ are shown in Figures \ref{f5} and \ref{f6}, respectively. Both show the hyperbolic structure of $\sigma$ around the interaction point. Thus the focusing of the beam till the interaction point has a trumpet structure, as shown in Figure \ref{f7}. The corresponding density plot of $|\psi|^2$ till the final focus in the transverse $x - y$ plane is shown in Figure \ref{f8}. We found the beam spot size at the final focus is $\sigma*\simeq 56$ nm.

\section{Conclusion and Remarks}\label{conclusion}
We have carried out a preliminary investigation to conceive a quantum plasma lens. We have considered the interaction of relativistic electron or positron beam with a magnetized plasma via PWF excitation by taking into account the quantum paraxial diffraction of the beam. The quantum regime for the electron or positron beam has been considered in such a way that we can take into account the individual quantum nature of the particle but disregarding the collective quantum behavior. Under these assumptions, the transverse effects experienced by the beam is investigated and the scheme of the plasma lens in the overdense regime has been reviewed.

In the present preliminary analysis, we have obtained a beam spot size at the final focus $\sigma*\simeq 56$ nm, which is three order smaller than the initial beam spot size.

We would like to point out thet the analysis has been carried out in the aberration-less approximation. Due to the feature of the medium (namely, the plasma), the linear focusing cannot be provided for all the values of the radial coordinate. In fact, distortions more or less important come into play during the focusing process. They cause the enhancement of the minimum beam spot size (aberrations). If the lens is conceived to be used at the final focusing stage of a linear collider, then such an enhancement of the minimum spot size causes the reduction of the luminosity at the interaction point.

In addition, collective effects on the classical and quantum scale such as beam-beam interaction and exchange interaction of a fermion system affect the luminosity. On the other hand, these two aspects come into play with special importance because our spot sizes are reduced to the nanoscales.

On the basis of the above considerations, a more careful analysis is under way. It takes into account both aberrations and collective effects related to both beamstrahlung and superposition of the single particle wave functions that we have disregarded in the present analysis.

Finally, we would like to emphasize that the transverse beam geometry encountered in conventional compact linear colliders differs from the cylindrically symmetric one that has been assumed here. This makes not easy the comparison of the two cases.

%\bibliographystyle{phjcp}
%% or: ieeetr,plain,unsrt,alpha,abbrv,acm,apalike,phjcp...
%\bibliography{Xbib}
\section{Acknowledgements}\nonumber
We thank Dr. Andrea Latina (CERN, Geneva, Switzerland) for valuable discussions concerning the features of CLIC (Compact Linear Collider).

\end{document}